\documentclass[twocolumn,aps,prd,superscriptaddress,nofootinbib,floatfix]{revtex4-1}

\usepackage{graphicx,multirow}
\usepackage{xspace}

\usepackage{hyperref}

\newcommand{\nn}{\nonumber}
\newcommand{\beq}{\begin{equation}}
\newcommand{\eeq}{\end{equation}}
\newcommand{\beqa}{\begin{eqnarray}}
\newcommand{\eeqa}{\end{eqnarray}}

\newcommand{\babar}{\mbox{\ensuremath{{\displaystyle B}\!{\scriptstyle A}{\displaystyle B}\!{\scriptstyle AR}}\xspace}}

\newcommand{\wmax}{w_{\rm max}}
\def\d{{\rm d}}
\newcommand{\Dprime}{\ensuremath{D^\prime}\xspace}
\newcommand{\Dprimestar}{\ensuremath{D^{\prime*}}\xspace}
\newcommand{\Dprimeboth}{\ensuremath{D^{\prime(*)}}\xspace}
\newcommand{\lqcd}{\ensuremath{\Lambda_{\rm QCD}}\xspace}

\newcommand{\Bbar}{\,\overline{\!B}{}}
\newcommand{\Dbar}{\,\overline{\!D}{}}
\newcommand{\Kbar}{\,\overline{\!K}{}}
\def\B0bar{\Bbar{}^0}
\def\D0bar{\Dbar{}^0}
\def\K0bar{\Kbar{}^0}

\begin{document}

\title{\boldmath A proposal to solve some puzzles in semileptonic $B$ decays}

\author{Florian U.\ Bernlochner}
\affiliation{University of Victoria, Victoria, British Columbia, Canada V8W 3P}

\author{Zoltan Ligeti}
\affiliation{Ernest Orlando Lawrence Berkeley National Laboratory,
University of California, Berkeley, CA 94720}

\author{Sascha Turczyk}
\affiliation{Ernest Orlando Lawrence Berkeley National Laboratory,
University of California, Berkeley, CA 94720}

\begin{abstract}

Some long-standing problems in the experimental data for semileptonic $b\to
c\ell\bar\nu$ decay rates have resisted attempts to resolve them, despite
substantial efforts.  We summarize the issues, and propose a possible
resolution, which may alleviate several of these tensions simultaneously,
including the ``1/2 vs.\ 3/2 puzzle" and the composition of the inclusive decay
rate in terms of exclusive channels. 

\end{abstract}

\maketitle

\section{Introduction}

There are several puzzling features of the semileptonic $b\to c$ decay data,
which have existed with varying level of significance for over ten years.  While
individually these are not many sigma problems, they affect several
measurements, and are a source of feeling uneasy about some semileptonic decay
results.  They relate to tensions between the measurements of inclusive and
exclusive decays.  The $D$ meson states relevant for our discussion are listed
in Table~\ref{tab:charm}. We refer to the first two states as $D^{(*)}$,
the next four as $D^{**}$, and the last two as $\Dprimeboth$.  The relevant
semileptonic decay measurements are~\cite{Nakamura:2010zzi, Asner:2010qj}:

\begin{enumerate}\vspace*{-6pt}\itemsep 0pt

\item The inclusive rate, ${\cal B}(B^+ \to X_c\ell^+\nu) = (10.92\pm 0.17)\%$,
and various inclusive spectra in this decay;

\item The exclusive rates ${\cal B}(B^+\to D\ell^+\nu) = (2.31 \pm 0.09)\,\%$
and  ${\cal B}(B^+\to D^*\ell^+\bar\nu) = (5.63 \pm 0.18)\,\%$;

\item The sum over the four rates, ${\cal B}(B^+\to D^{**} \ell^+ \nu) \times
{\cal B}(D^{**} \to D^{(*)}\pi) = (1.7 \pm 0.12)\,\%$, composed of roughly equal
rates for the sum over the two $s_l^{\pi_l} = \frac12^+$ and the two
$s_l^{\pi_l} = \frac32^+$ states;

\item The semi-inclusive rate ${\cal B}(B^+\to D^{(*)}\pi\ell^+\nu) = (1.53 \pm
0.13)\,\%$, including a $D^{(*)}$ and exactly one $\pi$.

\end{enumerate}\vspace*{-6pt}

The sum of the measured exclusive rates is less than the inclusive one (the
value in item 1.\ is obtained from the more precise average branching ratio for
$B^0$ and $B^\pm$ using equal semileptonic rates), and  previous attempts to
bring the two into agreement have faced problems.  In particular, the inclusive
rate (1.) minus those in items 2.\ and 4.\ gives $(1.45 \pm 0.29)\%$.  Assigning
this to semileptonic $B$ decays to other nonresonant channels (i.e., to final
states containing more than one hadrons, not included in item 4.), results in a
too soft inclusive charged lepton energy spectrum, inconsistent with the data. 
There has also been a persistent $\sim2\sigma$ difference between $|V_{cb}|$
extracted from inclusive and exclusive semileptonic $B$ decays.

The charm meson states relevant for our discussion are organized as doublets of
heavy quark spin symmetry, and are shown in Table~\ref{tab:charm}. The $D$ and
$D^*$ states are the $1S$ ground state in the quark model. The next four
$D^{**}$ states are the $1P$ orbital excitations (with the spin and parity of
the brown muck equal $s_l^{\pi_l} = \frac12^+$ and $\frac32^+$), and the \Dprime
and \Dprimestar states correspond to the first radial excitation in the quark
model (the $2S$ states).

Another issue which has received a lot of attention, but concerns a tension not
simply between different pieces of data, but the comparison of theory with data,
is the ``1/2 vs.\ 3/2 puzzle".  Model calculations predict that semileptonic $B$
decays should have a substantially smaller rate to the $s_l^{\pi_l} = \frac12^+$
doublet than to the $s_l^{\pi_l} = \frac32^+$ doublet~\cite{Morenas:1997nk,
Bigi:2007qp}, contrary to what is observed (item 3.\ above).

\begin{table}[b]
\begin{tabular}{ccccc}
\hline\hline
Notation  &  $s_l^{\pi_l}$ &  $J^P$  &  $m$ (GeV)  &  $\Gamma$ (GeV)\\
\hline
$D$  &  $\frac12^-$  &  $0^-$ &  $1.87$  \\
$D^*$  &  $\frac12^-$  &  $1^-$  &  $2.01$  \\[2pt] \hline
$D_0^*$  &  $\frac12^+$  &  $0^+$  &  $2.40$  &  0.28 \\
$D_1^*$  &  $\frac12^+$  &  $1^+$  &  $2.44$  & 0.38  \\ \hline
$D_1$  &  $\frac32^+$  &  $1^+$  &  $2.42$  & 0.03 \\
$D_2^*$  &  $\frac32^+$  &  $2^+$  &  $2.46$  &  0.04 \\ \hline
$D'$  &  $\frac12^-$  &  $0^-$ &  2.54  & 0.13 \\
$\Dprimestar$  &  $\frac12^-$  &  $1^-$  &  2.61  &  0.09  \\
\hline\hline
\end{tabular}
\caption{Charm meson states and their isospin averaged masses and widths.
\Dprimeboth denote the $2S$ excitation of $\Dprimeboth$.
The $s_l^{\pi_l}$ is the spin and parity of the light degrees of freedom,
which is a good quantum number in the heavy quark limit~\cite{Isgur:1991wq}.}
\label{tab:charm}
\end{table}

In past experimental analyses there have been various approaches to deal with
these issues, typically making cuts with the hope of eliminating the effects of
these discrepancies, and/or modifying some of the exclusive rates in the event
generators.

Here we propose that these problems could be eased (or maybe even solved) by an
unexpectedly large $B$ decay rate to the first radially excited \Dprimeboth
states.  Recently \babar\, found evidence for two new
states~\cite{delAmoSanchez:2010vq}, which are most likely these $2S$ states in
the quark model picture~\cite{Godfrey:1985xj}.

\section{Proposal and viability}

We would like to explore the possibility that the sum of the two \Dprimeboth
decay rates is substantial,
\beq\label{postulate}
{\cal B}\big(B\to \Dprimeboth\ell\bar\nu\big) \sim {\cal O}(1\,\%)\,, 
\eeq
and show that it can help resolve the problems mentioned above, without
giving rise to new ones.

\begin{figure*}[tb]
   \includegraphics[width=\columnwidth]{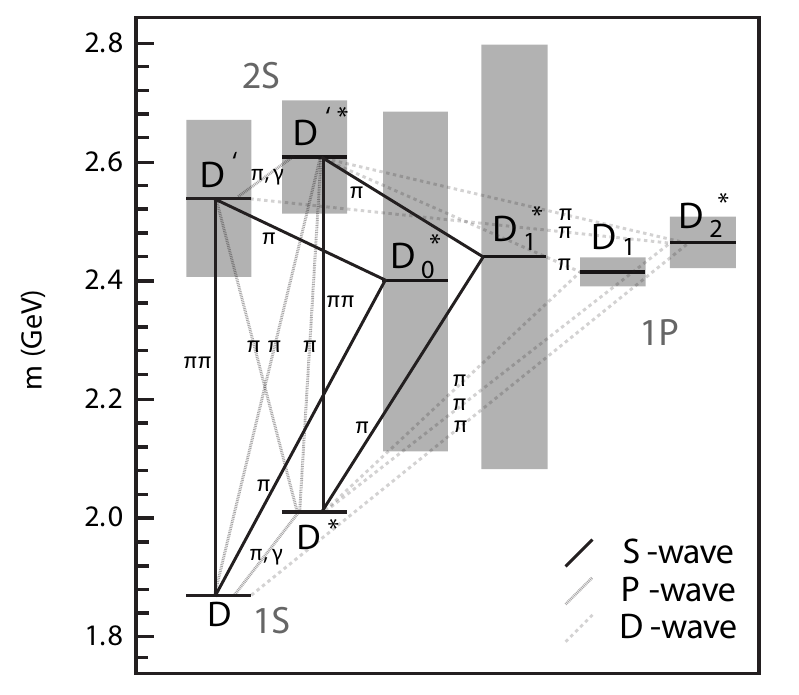} \hfill
   \includegraphics[width=\columnwidth]{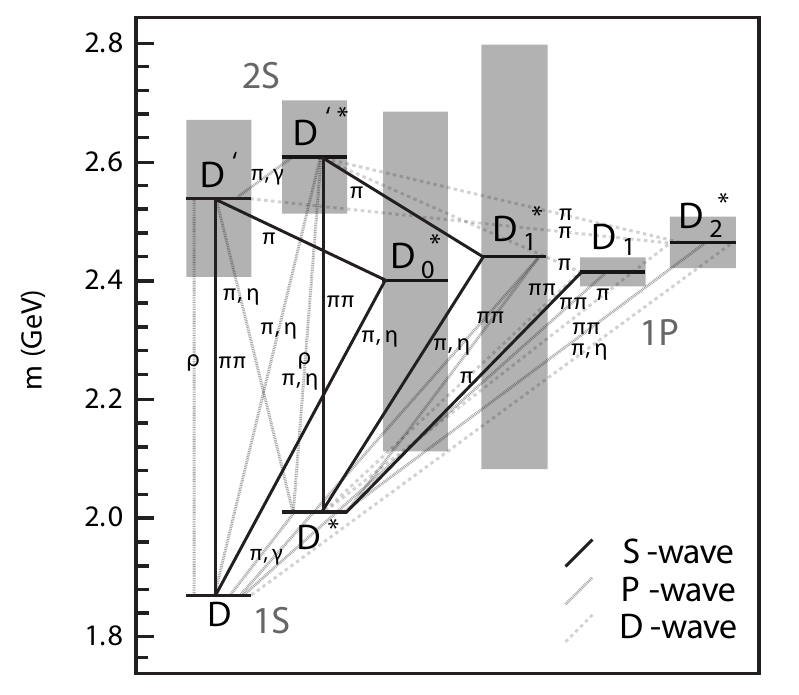}
   \caption{Strong decays of the $D'$ and $\Dprimestar$ into the $1S$ and $1P$
   states involving, one or two pion emissions (left), and all decays including
   the near off-shell transitions with a $\rho$ and $\eta$ (right). The style
   and opacity of the lines connecting the states indicate the orbital angular
   momentum of the partial wave. The grey bands correspond to the measured
   widths of the $2S$ and $1P$ states.}
   \label{ill:charmall}
\end{figure*}

1) The rate in Eq.~(\ref{postulate}) would be a big enough contribution to the
sum over exclusive states, so that the nonresonant
contribution~\cite{Goity:1994xn} no longer needs to be large.  This would be a
problem, because in the soft pion limit a first principles calculation is
possible~\cite{Wise:1992hn}, giving a too small rate at this region of phase
space.  A large nonresonant rate at high $D^{(*)}\pi$ invariant mass would
disagree with the inclusive lepton spectrum measurements and the measured
semi-exclusive $B\to D^{(*)}\pi\ell\bar\nu$ rate. 

2) The $\Dprimeboth$ states decay to one of the $D^{(*)}$ states either with one
pion emission in a $p$-wave, or with two pion emission in an $s$-wave.  However,
they can decay with one pion emission in an $s$-wave to members of the
$s_l^{\pi_l} = \frac12^+$ states, and could thus enhance the observed decay rate
to the $s_l^{\pi_l} = \frac12^+$ states, and thus give rise to the ``1/2 vs.\
3/2 puzzle". The allowed strong decays are illustrated in
Figure~\ref{ill:charmall} (including those only allowed by the substantial
widths of these particles). It is plausible that the decay modes of the
\Dprimeboth to the $1S$ and $1P$ charm meson states may be comparable. 

3) With the relatively low mass of the \Dprimeboth states, the inclusive
lepton spectrum can stay quite hard, in agreement with the observations.

4) The ${\cal B}(B\to D^{(*)}\pi\ell\bar\nu)$ measurement quoted is not in
conflict with our hypothesis, since the decay of the \Dprimeboth would yield two
or more pions most of the time.

\section{\boldmath The $B\to \Dprimeboth\ell\bar\nu$ decay rate}

Since the quantum numbers of the \Dprimeboth are the same as those of the
$D^{(*)}$, the theoretical expressions for the decay rates in terms of the form
factors, and the definitions of the form factors themselves, are identical to
the well known formulae for $B\to D^{(*)}\ell\bar\nu$~\cite{Manohar:2000dt}.  As
for $B\to D^{(*)}\ell\bar\nu$, in the $m_{c,b} \gg \lqcd$ limit, the six form
factors are determined by a single universal Isgur-Wise
function~\cite{Isgur:1989vq}, which we denote by $\xi_2(w)$.  Here $w=v \cdot
v'$ is the recoil parameter, $v$ is the velocity of the $B$ meson, and $v'$ is
that of the \Dprimeboth.  We define
\beqa\label{rates}
{{\rm d}\Gamma_{\Dprimestar}\over {\rm d}w} &=& 
  \frac{G_F^2 |V_{cb}|^2\, m_B^5}{48\pi^3}\, r^3 (1-r)^2\, \sqrt{w^2-1}\,
  (w+1)^2 \nn\\*
&&{} \times \bigg[ 1 + \frac{4w}{w+1}\, \frac{1-2rw+r^2}{(1-r)^2} \bigg] 
  \big[F(w)\big]^2\,, \\[4pt]
{{\rm d}\Gamma_{\Dprime}\over {\rm d}w} &=& 
  \frac{G_F^2 |V_{cb}|^2\, m_B^5}{48\pi^3}\, r^3 (1+r)^2\, (w^2-1)^{3/2}\,
  \big[G(w)\big]^2\,, \nn
\eeqa
where, in each equation, $r = m_{\Dprimeboth}/m_B$, and in the $m_{c,b} \gg
\lqcd$ limit $F(w) = G(w) = \xi_2(w)$.

Heavy quark symmetry implies $\xi_2(1)=0$, so the rate near zero recoil comes
entirely from $\lqcd/m_{c,b}$ corrections.  Away from $w=1$, $\xi_2(w)$ is no
longer power suppressed; however, since the kinematic range is only $1 < w <
1.3$, the role of $\lqcd/m_{c,b}$ corrections, which are no longer universal,
can be very large~\cite{Leibovich:1997em}.  Before turning to model
calculations, note that there is a qualitative argument that near $w=1$ the
slope of $\xi_2(w)$, and probably those of $F(w)$ and $G(w)$ as well, should be
positive.  In $B\to \Dprimeboth$ transition, in the quark model, the main effect
of the wave function of the brown muck changing from the $1S$ to the $2S$ state
is to increase the expectation value of the distance from the heavy quark of a
spherically symmetric wave function.  Thus the overlap of the initial and final
state wave functions should increase as $w$ increases above 1.

\begin{figure*}[bt]
   \includegraphics[width=\columnwidth]{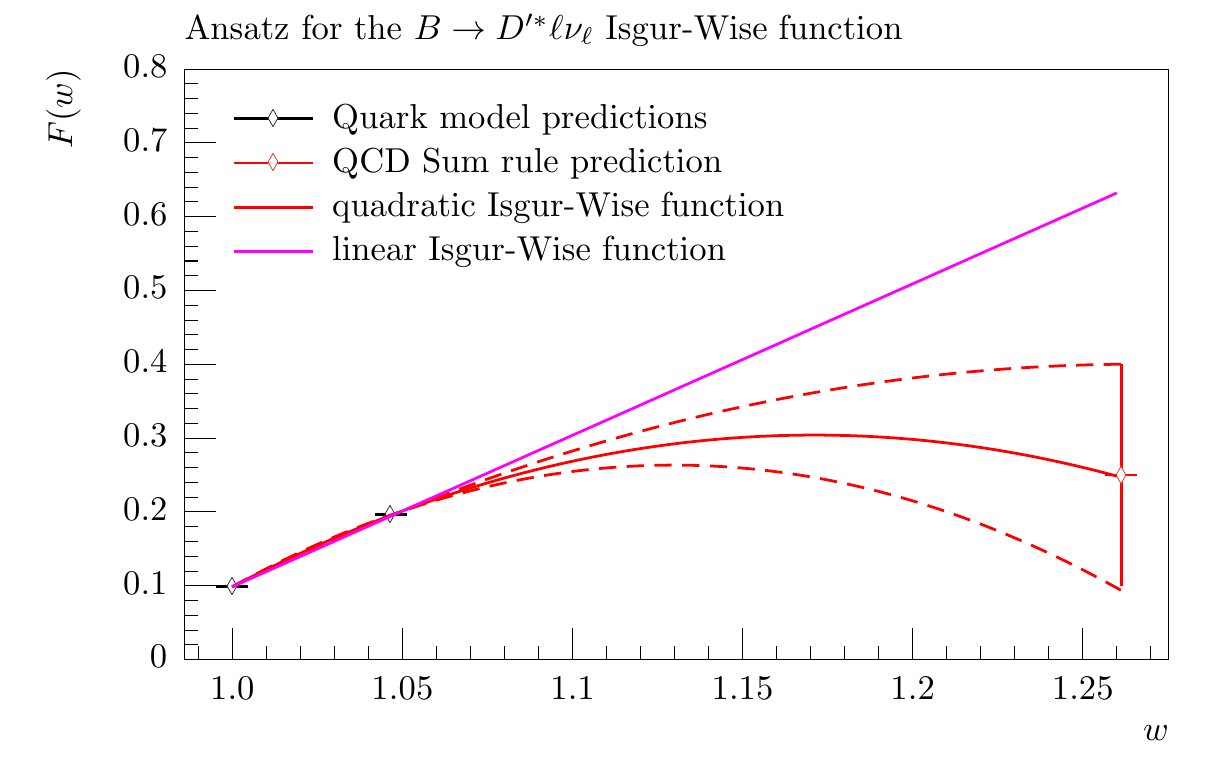} \hfill
   \includegraphics[width=\columnwidth]{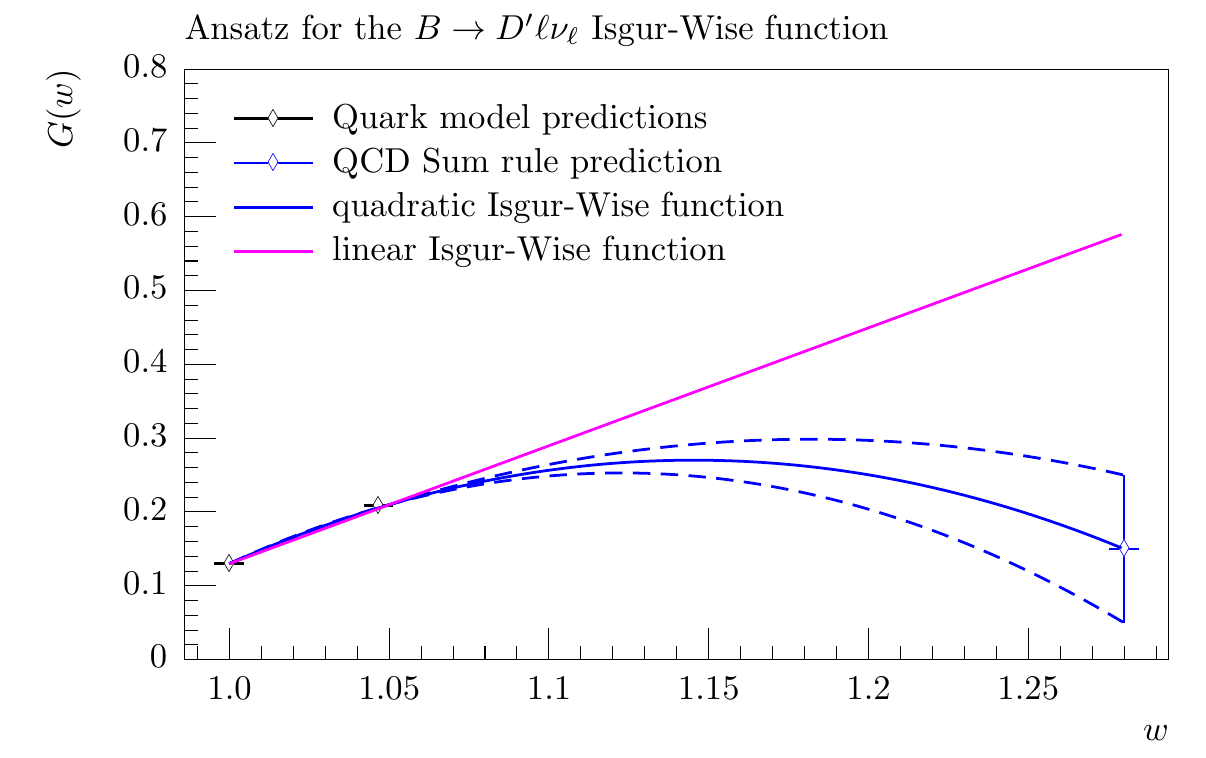}
   \caption{The function $F(w)$ which determines the $B\to
   \Dprimestar\ell\bar\nu$ rate (left) and $G(w)$ which determines $B\to
   \Dprime\ell\bar\nu$ (right).
   The quark model calculations at $w=1$ and $1.05$ are compared with the sum
   rule prediction at $\wmax$. The solid lines show a quadratic and linear ansatz for the
   Isgur-Wise function and the dashed lines correspond to the variation of the
   sum rule parameters.}\label{iwfn}
\end{figure*}

It is not easy to calculate these $B\to \Dprimeboth\ell\bar\nu$ form factors. 
Below, we use estimates from a quark model prediction~\cite{Ebert:1999ed}, hoped
to be trustable near $w = 1$, and from modifying a QCD light-cone sum rule
calculation~\cite{Faller:2008tr}, hoped to be reasonable near maximal recoil. 
We emphasize that both models were developed, tuned, and tested for states that
are the lightest with a given set of quantum numbers.  Thus, one should take the
following numerical estimates with a truck load of salt, and we present them
only to substantiate that the rate in Eq.~(\ref{postulate}) is plausible and
should be searched for experimentally.  The same physical problem (and the width
of the $\Dprimeboth$) would also provide a formidable challenge to lattice QCD
calculations of the  $B\to \Dprimeboth\ell\bar\nu$ form factors. 

A quark model calculation of the values and slopes of the leading and subleading
Isgur-Wise functions at zero recoil~\cite{Ebert:1999ed}, imply
\beqa\label{quarkmodel}
    F(1.0) &=& 0.10\,,\qquad  F(1.05) = 0.20 \,, \nn\\*
    G(1.0) &=& 0.13\,,\qquad  G(1.05) = 0.21 \,,
\eeqa
where the values at $w=1.05$ are obtained from a linear extrapolation.
Perturbative QCD corrections enhance $F'(1)$ compared to $G'(1)$, by about
0.1~\cite{Grinstein:2001yg}.

The light-cone sum rule calculation~\cite{Faller:2008tr} can in principle be
adapted to extract the $B\rightarrow \Dprimeboth $ form factors, i.e., the first
radial excitation, near maximal recoil.  We obtain the following estimates (the
technical details are described in the Appendix), 
\beq\label{srres}
    F(\wmax) = 0.25 \pm 0.15\,, \qquad  G(\wmax) = 0.15 \pm 0.1\,. 
\eeq
As one may anticipate, the largest uncertainty originates from the way the 
suppression of the ground state $D^{(*)}$ contribution is achieved, to project
out the first radially excited state from the hadronic dispersion relation.

We parametrize the $F(w)$ and $G(w)$ functions which determine the $\Dprimeboth$
decay rates as quadratic polynomials, which is sufficient for our purposes,
\beqa\label{eq:IWfunc}
  F(w) &=& \beta_0^* + (w-1) \beta_1^* + (w-1)^2 \beta_2^*\,, \nn\\*
  G(w) &=& \beta_0 + (w-1) \beta_1 + (w-1)^2 \beta_2 \,.
\eeqa
To get a rough estimate of the possible $B \to \Dprimeboth \ell\bar\nu$ rates,
we determine the parameters in Eq.~(\ref{eq:IWfunc}) to predict $F(w)$ and
$G(w)$, as shown in Figure~\ref{iwfn}.  Using the simple quadratic
parametrization in Eq.~(\ref{eq:IWfunc}) together with Eqs.~(\ref{quarkmodel})
and (\ref{srres}) yield for the branching fraction of the sum of the two
semileptonic $B \to \Dprimeboth \ell \nu_\ell$ decays
\beq
 \mathcal{B}\big(B \to \Dprimeboth \ell \nu_\ell\big) \sim (0.3-0.7)\,\% \,,
\eeq
with the parameters
\beqa
 \beta_0^* &=& 0.10 \,,\quad \beta_1^* = 2.3-2.5 \,,\quad \beta_2^* = -(4.2-9.8)
 \,, \nn \\
 \beta_0 &=& 0.13 \,,\quad \beta_1 = 1.9-2.0 \,,\quad \beta_2 = -(5.1-8.2) \,.
\eeqa
Earlier quark model calculations, without accounting for $\lqcd/m_{c,b}$
effects, obtained smaller rates~\cite{Scora:1995ty, Suzuki:1993iq}, while
including $\lqcd/m_{c,b}$ effects, 0.4\% was obtained~\cite{Ebert:1999ed}.
If, instead, we use a linear parametrization and the quark model results in 
Eq.~(\ref{quarkmodel}) only, then we get
\beq
 \mathcal{B}\big(B \to \Dprimeboth \ell \nu_\ell\big) \sim 1.4 \% \, ,
\eeq
with the parameters
\beqa
  \beta_0^* &=& 0.10 \,,\qquad \beta_1^* = 2.1 \, , \nonumber \\
  \beta_0 &=& 0.13 \,,\qquad \beta_1 = 1.6 \, .
\eeqa
We take these as indications that Eq.~(\ref{postulate}) is plausible, and $B \to
\Dprimeboth \ell\bar\nu$ may account for a substantial part of the observed
``gap" between inclusive and exclusive decays.

Another measurement that can constrain this picture are the nonleptonic rates,
$B\to \Dprimeboth\pi$. Factorization, which was proven to leading order in the
heavy mass limit in the decays we consider~\cite{Bauer:2001cu}, implies that in
each channel the nonleptonic rate is related to the semileptonic differential
decay rate at maximal recoil,
\beq\label{Lbfactor2}
\Gamma(B\to \Dprimeboth\pi) = {3\pi^2 C^2\, |V_{ud}|^2 f_\pi^2
  \over m_B\, m_{\Dprimeboth}}\, {\d\Gamma(B\to
  \Dprimeboth\ell\bar\nu)\over \d w} \Bigg|_{\wmax} .
\eeq
Here $C$ is a combination of Wilson coefficients and numerically $C\, |V_{ud}|
\approx 1$, and $\wmax$ corresponds to $q^2 = 0 \simeq m_\pi^2$. Thus, besides a
direct search for $B\to \Dprimeboth \ell\bar\nu$ decays, measuring the
nonleptonic $B\to \Dprimeboth\pi$ rates would also be very valuable to
constrain $F(w)$ and $G(w)$.  This type of measurement, including a Dalitz plot
analysis of $\Bbar \to [D^{(*)} \pi^+ \pi^-] \pi^-$, would also be valuable in
understanding the decay rates of the \Dprimestar states.

\section{conclusions}

If future measurements find a substantial $B \to \Dprimeboth \ell \bar \nu$
decay rate, the precise determination of the branching fraction, the shape of
the $F(w)$ and $G(w)$ functions in Eq.~(\ref{rates}), and data on the
corresponding nonleptonic two-body decays with a pion would be able to test this
picture.  It may also impact other measurements and the theory of semileptonic
decays, e.g., it may yield

\begin{itemize}\vspace*{-4pt}\itemsep 0pt

\item a better understanding of the $b \to c$ background in fully inclusive
$b \to u$ measurements, i.e., lead to a more precise determination of $|
V_{ub}|$;

\item a better understanding of the semileptonic $b \to c$ background in the
exclusive $| V_{cb}|$ measurements using $B \to D^{(*)} \ell\bar\nu$;

\item a better understanding of the missing exclusive contributions to the
inclusive $B\to X_c\ell\bar\nu$ rate, and the lepton energy and hadronic mass
spectrum;

\item a better understanding of the measured $B \to D^{(*)} \tau \bar \nu$ branching fraction 
and its tension with respect to the Standard Model expectation \cite{FrancoSevilla:2011};

\item a more precise determination of the semileptonic branching fractions of
the $s_l^{\pi_l} = \frac12^+$  and $\frac32^+$ states, thus maybe help resolve
the ``1/2 vs.\ 3/2 puzzle";

\item a stronger sum rule bound~[\onlinecite{Bigi:1994ga, Kapustin:1996dy,
Dorsten:2003ru}, \onlinecite{Leibovich:1997em}] on the $B\to D^*\ell\bar\nu$
form factor, ${\cal F}(1)$, relevant for the determination of $|V_{cb}|$ from
exclusive decay.

\end{itemize}\vspace*{-4pt}

There are a number of measurements that should be possible using the \babar,
Belle, LHCb, and future $e^+e^-$ $B$ factory data samples, which could shed
light on whether this possibility is realized in nature.

\begin{acknowledgments}

We thank Alexander Khodjamirian, Christoph Klein, Bob Kowalewski,  Heiko Lacker,
and Vera L\"uth for helpful discussions.  FB thanks the LBNL theory group for
their hospitality.  This work was supported in part by the Director, Office of
Science, Office of High Energy Physics of the U.S.\ Department of Energy under
contract DE-AC02-05CH11231. ST~is supported by a DFG Forschungsstipendium under
contract no.~TU350/1-1. 

\end{acknowledgments}

\appendix*
\section{Details of the Sum Rule Estimate}

The crucial ingredient in obtaining Eq.~(\ref{srres}) is to modify the
light-cone sum rule (LCSR) computation~\cite{Faller:2008tr} in a manner that the
radially excited state can be projected out.  We are not aware of similar
attempts, so we give some details of our calculation.  We write out explicitly
the pole of the radial excitation in the hadronic dispersion relation and
multiply the formula by the ground state pole, e.g., schematically shown for the
decay constant
\beq
    \frac{m_D^4 f_D^2}{m_c^2(m_D^2- q^2)} +
    \frac{m_{\Dprime}^4 f_{\Dprime}^2}{m_c^2 (m_{\Dprime}^2- q^2)} + 
    \int_{s_0^{\Dprime}}^\infty \text{d}s \frac{\rho(s)}{s-q^2}\,.
\eeq
Projecting out the radial excited state amounts to modifying the Borel
transformation according to 
\beq
    B_{q^2} \frac{m_D^2-q^2}{(s-q^2)^k} = 
    \frac{\left[(k-1) m^2 - (s-m_D^2)\right]}{(k-1)!} 
    \frac{e^{-s/m^2}}{(m^2)^{k-1}}\,,
\eeq
which leads to a correction term for the sum rule for the form factors of the
first radially excited state compared to the expressions
in~\cite{Faller:2008tr}. In order to use the known LCSR up to three particle
contributions, one needs to apply a correction term to the Borel transform,
$\zeta_k(s,m^2,m_D^2, m_{\Dprime}^2)$ for $k=1,2,3$ 
\beq
    \zeta_k(s,m^2,m_D^2, m_{\Dprime}^2) = 
    \frac{(k-1) m^2 - (s-m_D^2)}{m_D^2-m_{\Dprime}^2}\,.
\eeq
Due to necessary partial integration, it is a nontrivial endeavor to implement
this correction term. A more detailed study of the numerical stability is, in
principle, possible by multiplying the formulae with higher powers of the ground
state pole. This modifies the correction term $\zeta_k$ according to
\beq
(m_D^2-q^2)^n = \sum_{k=0}^n \Big(\begin{array}{c} n \\ k \end{array}\Big)
(-q^2)^{k} (m_D^2)^{n-k}\,, 
\eeq
but does not affect the formal ground state suppression, since $B_{q^2} (q^2)^k
= 0$ for $k\geq 0$. However, this is beyond the scope of our estimate in the
context of this analysis, so we quote the results for the simplest calculation.

The resulting values for the form factors are sensitive to
the numerical input values for the decay constants, the Borel and duality
parameters. The latter parameters can be varied to estimate the sensitivity of
the final result. The duality parameter, which has to be chosen higher than the
corresponding meson mass, approximates the spectral density over the remaining
physical resonances.\footnote{For the $1S$ state this parameter can be estimated
by demanding to reproduce the meson mass, which is not possible for the $2S$.}
Presumably for higher excited states the ratio $s_0^D/m_D^2$ of the
corresponding state should be chosen higher than usual, due to the spectral
density shape. The Borel parameter $m$, which suppresses exponentially the
higher states, needs to be chosen large enough to obtain a reliable perturbative
expansion, but small enough to not loose the sensitivity to the radially excited
state (the influence of higher dimensional quark condensates increases with
decreasing Borel parameter). Compared to Ref.~\cite{Faller:2008tr}, additional
uncertainties emerge from (i)~the approximate suppression of the ground state;
(ii)~the smaller separation to higher excited states; and (iii)~larger
perturbative and nonperturbative corrections. A further complication arises due
to the poor knowledge of the $\Dprimeboth$ decay constants, which are needed as
an input to the sum rules. Following a similar approach as in
Ref.~\cite{Jamin:2001fw}, we estimate the decay constants of the radially
excited states, which prove fairly sensitive to the particular choice of Borel
and duality parameters. We assume that the ratio of the decay constants for the
radial excited states should be similar to that in the ground state, i.e.,
$f_{\Dprimestar}/f_{D^{*}} \sim 1.4$, which holds for the parameters we choose,
\beq
    f_{\Dprimestar} \sim 300\,\text{MeV} \,,\qquad
    f_{\Dprime} \sim 200\,\text{MeV}\,.
\eeq
The $\Dprimeboth$ decay constants enter the sum rules, and are an additional
source of uncertainty. We find a stable plateau for the various form factors
with respect to the Borel and duality parameters, yet at values which should be 
too high from physical considerations. For the quoted $\Dprime$ and
$\Dprimestar$ form factors we choose a duality parameter of $s_0^{\Dprime} =
15\,\text{GeV}^2$ and $s_0^{\Dprimestar} = 17\,\text{GeV}^2$, respectively, and
a common Borel parameter of $m^2 = 7 \,\text{GeV}^2$, which are smaller than the
ones at the plateau, resulting in a smaller form factor. The parameters chosen
for computing the ground state yields a value close to zero for both form
factors, in agreement with the expected suppression of the ground state
contribution.

One may be concerned about the level of heavy quark symmetry violation, such as
the deviation of $F(w)/G(w)$ from unity. Deviations are due to $\lqcd/m_{b,c}$
effects as well as perturbative corrections. Using the sum rule prediction,
one obtains
\beq
    F(\wmax) \big/ G(\wmax) = 1.7 \pm 0.6\,.
\eeq
where we assumed a 90\% correlation between the uncertainties due to the choice
of the Borel and duality parameters. For the form factor ratios $R_1$ and $R_2$
we obtain at maximal recoil, 
\beq
   R_1(\wmax)= 2.0 \pm 0.4 \,,\qquad
   R_2(\wmax)= 1.1 \pm 0.3 \,.
\eeq
Interestingly, their ratio, $R_1(\wmax) /R_2(\wmax) = 1.8 \pm 0.2$, is
not far from the similar ratio for the $D^*$ case.

\end{document}